\documentclass[aps,prb,twocolumn,superscriptaddress,showkeys]{revtex4-2}
\usepackage{amsmath,amssymb}
\usepackage{graphicx}
\usepackage{dcolumn}
\usepackage{bm}
\usepackage{esvect}
\usepackage{float}
\usepackage{color}
\usepackage{amsmath}
\usepackage{amsfonts}
\definecolor{rojo}{rgb}{1,0,0}
\definecolor{verde}{rgb}{0,0.8,0.2}
\definecolor{azul}{rgb}{0,0,1}
\definecolor{rosa}{cmyk}{0,1,0,0}
\usepackage[toc,page]{appendix} 
\usepackage[export]{adjustbox}
\usepackage{latexsym}
\usepackage{mathrsfs}
\usepackage{bm}
\usepackage{physics}
\usepackage{amsthm}
\usepackage{bbm}
\usepackage{amsfonts}
\usepackage{color}
\usepackage{epsfig}
\usepackage{hyperref}
\hypersetup{colorlinks=true, 
            citecolor=blue,
            urlcolor=black,
            linkcolor=blue}
\usepackage{soul}
\DeclareGraphicsExtensions{.pdf,.png,.jpg,.pdf}
\usepackage{multirow}
\usepackage{changes}
\usepackage{fancyvrb}
\newcommand{\mrs}{Mn$_2$RhSn}
\DeclareUnicodeCharacter{0301}{\'{e}}

\begin{document}

\title{Exploring non-collinear magnetic ground states \\ in tetragonal Mn$_2$-based Heusler compounds }

\author{Jorge Cardenas-Gamboa}
\email{jorge.cardenas@cpfs.mpg.de}
\affiliation{Max Planck Institute for Chemical Physics of Solids, 01187 Dresden, Germany}

\author{Edouard Lesne}
\affiliation{Max Planck Institute for Chemical Physics of Solids, 01187 Dresden, Germany}

\author{Arthur Ernst}
\affiliation{ Institute for Theoretical Physics, Johannes Kepler University of Linz, Altenberger Strasse 69, A-4040 Linz, Austria.}
\affiliation{Max-Planck-Institut f\"ur Mikrostrukturphysik, Weinberg 2, 06120 Halle (Saale), Germany}

\author{Maia G. Vergniory}
\email{maia.vergniory@usherbrooke.ca}
\affiliation{ Donostia International Physics Center, Donostia-San Sebastian 20018 Gipuzkoa, Spain}
\affiliation{D\'epartement de physique et Institut quantique, Universit\'e de Sherbrooke, Sherbrooke J1K 2R1 QC, Canada}
\affiliation{Regroupement Qu\'eb\'ecois sur les Mat\'eriaux de Pointe (RQMP), Quebec H3T 3J7, Canada}

\author{Paul McClarty}
\email{paul.mcclarty@cea.fr}
\affiliation{ Laboratoire Le\'on Brillouin, CEA, CNRS, Universite ́ Paris-Saclay, CEA Saclay, 91191 Gif-sur-Yvette, France}

\author{Claudia Felser}
\affiliation{Max Planck Institute for Chemical Physics of Solids, 01187 Dresden, Germany}

\date{\today}

\begin{abstract}
Heusler compounds constitute a large family of intermetallic materials notable for their wide variety of properties such as magnetism, multi-ferroicity, nontrivial band topology, superconductivity and so on. Among their magnetic properties one finds a tremendous variety of states from simple ferromagnetism to skyrmion crystals. In most Mn$_2$-based Heuslers the magnetism is typically collinear. An exception is \mrs\ in which an unusual ground state with magnetic canting and a temperature-induced spin re-orientation into the collinear ferrimagnetic phase has been reported from experiments. In this work, we employ first-principles calculations and mean field theory to provide a simple account of the unusual phase diagram in this magnet. We also highlight Weyl points in the computed band structure of \mrs\ and the resulting Fermi arcs.
\end{abstract}

\maketitle
\section{Introduction}

It is of both fundamental and practical interest to investigate magnetism in metallic systems, whereby intertwined magnetic and itinerant degrees of freedom give rise to a plenty of magnetic ground states and spin-dependent electronic transport properties. The field of spintronics, which tackles such interplay and aims at practical applications in \textit{e.g.}, magnetic recording or random access magnetic memory devices, has however largely been focused on collinear magnetic systems ~\cite{intro_hirohata2020review}. However, as of relatively recently non-collinear magnetic states, whether low temperature ground states or low-energy non-collinear spin textures with magnetic Skyrmions being the prime example, are now at the center of attention in solid state physics and spintronics communities ~\cite{intro_fert2017magnetic,intro_tokura2020magnetic}.


Heusler compounds are a remarkable class of intermetallic materials that have garnered significant interest over the past decades due to their wide range of exceptional properties ~\cite{heusler1_graf2011simple,heusler2_wollmann2017heusler}. These materials are of particular interest for their magnetic and topological features. One of the most fascinating aspects of Heusler compounds is their versatility in accommodating a broad variety of magnetic ground states including ferromagnetic (FM), antiferromagnetic (AFM), ferrimagnetic (FiM), and non-collinear spin structures, depending on their chemical composition~\cite{heusler_5_palmstrom2016heusler}. 

For instance, Co$_2$MnSi is a prototypical ferromagnetic Heusler compound with high spin polarization, making it a benchmark material for spintronics ~\cite{heusler_3_co2MnSipicozzi2004role}. From a transport perspective, Heusler compounds exhibit exceptional properties, with Co$_2$MnGa standing out as a magnetic Weyl semimetal that notably exhibits a large anomalous Hall effect (AHE) and robust transport behavior at room temperature ~\cite{heusler_7_co2mnga_ludbrook2017perpendicular,weyl_guin2019zero}. Hexagonal Mn$_3$Sn and Mn$_3$Ge are remarkable examples of non-collinear antiferromagnetic compounds among the Heusler family, which extensively studied due to their large anomalous Hall effect driven by topologically nontrivial band structures~\cite{heusler_4_mn3Sn,hall_kubler2014non}. Such materials not only bridge the gap between magnetism and topology but also open doors to applications in energy-efficient devices and next-generation quantum technologies.

In this paper we consider magnetism in tetragonal inverse Heuslers ~\cite{felser2015basics,manna2018heusler} with chemical formula Mn$_2$XZ where X is a transition metal, Z is a main-group element and where the manganese ions are the only ones carrying a magnetic moment. This family of materials mainly exhibits non-centrosymmetric crystal structures. Many also have collinear magnetism enriched by the fact that two distinct Wyckoff positions in the crystal are occupied by manganese making ferrimagnetism natural in these systems. Among the magnetism in this class of materials, two stand out as their magnetic ground states are non-collinear. These are the half-metallic itinerant magnets Mn$_2$RhSn ~\cite{meshcheriakova2014large} and Mn$_2$IrSn ~\cite{mn_ir_sn_mozur2024magnetic}. In the original experimental work of Ref.~~\cite{meshcheriakova2014large} on \mrs\ it was argued that a competition between exchange couplings could lead to canting of the moments in the ground state. Here we expand on that study providing a minimal exchange model, which accounts both for the non-collinearity in the ground state and the reorientation of the moments at finite temperature. A crucial point of this model is the observation that the two shortest range exchange couplings can lead to collinear antiferromagnetism and to a spiral phase, whereas the exchange to third neighbor (with two symmetry-distinct couplings) can destabilize the antiferromagnetism leading to the observed canted structure. Additionally, we perform first-principles calculations to derive the exchange interactions and assess their influence on the magnetic properties, extending our analysis to other tetragonal inverse Heusler magnets. We then explore the topological properties that arise from the magnetism focusing on the Weyl nodes and surface states.

This work is organized as follows: First, we describe the main experimental features of \mrs. Then, we set up a minimal exchange model and show that it accounts for main features of the magnetism in Mn$_2$RhSn and Mn$_2$IrSn. Next, we validate these interactions using an \textit{ab initio} model, from which we extract the exchange coupling parameters to compare with our model. We then analyze the Weyl points and Fermi arcs in \mrs. Finally, we conclude by summarizing the key findings of this study. 

\section{Experimental results}

We study the inverse tetragonal Mn$_2$-based Heusler compounds, which crystallize in the non-centrosymmetric $I\bar{4}m2$ space group (No. 119). Experimental research (see Ref.~\cite{meshcheriakova2014large}) reports that the Mn atoms occupy two inequivalent Wyckoff positions: Mn$_I$ atoms are located at 2b (0,0,1/2), while Mn$_{II}$ atoms are positioned at 2c (0,1/2,3/4). Additionally, Rh and Sn atoms occupy the 2d and 2a Wyckoff positions, respectively. In Heusler compounds, strong coupling between nearest-neighbor Mn moments ($\it{J_{Mn_{I}-Mn_{II}}} \sim -$20~meV) typically results in a collinear ferrimagnetic ordering ~\cite{fim_col_meinert2011exchange}. However, it has been demonstrated that the inclusion of nonmagnetic heavy elements, such as Rh and Ir, induces spin reorientation accompanied by the emergence of a non-collinear magnetic order ~\cite{wollmann2015magnetism, faleev2017origin}. Consequently, translational symmetry is broken, leading to the formation of distinct magnetic sublattices (see Fig.~\ref{figure1}(a)) with propagation vector $k=(\frac{1}{2},\frac{1}{2},\frac{1}{2})$. 

Neutron scattering data obtained at the lowest measured temperature ($T = 1.8$ K) reveal a canted non-collinear magnetic structure with an angle of $\theta = (180 \pm 58.9)^\circ$ within alternating Mn$_{II}$-Rh planes.  Observation of the (002) magnetic diffraction peak for $T \leq 80$ K the presence of an in-plane moment component. Below 80 K, in-plane magnetism begins to rapidly evolve, driven by the deviation of the Mn$_{II}$ moment from axis. In contrast, the moment of Mn$_{I}$ remains strongly localized and firmly aligned along the c-axis (see Fig.~\ref{figure1}(b)). This in-plane magnetism arises from the gradual spin reorientation of the Mn$_{II}$ sublattice, involving changes not only in the mutual orientation of site-specific moments but also in their absolute values. Temperature-dependent measurements reveal a transition from a non-collinear magnetic phase at the lowest investigated temperature to a ferrimagnetic (FiM) phase around 80 K. The vanishing magnetization components when approaching 280 K corresponds to the Curie temperature ($T_\text{C}$) (see Fig.~\ref{figure1}(c)).

\section{Mean-Field model} \label{mean-field}

To understand the origin of this magnetic phases, a mean-field approximation is a functional method accounting for the competing interactions within. We have implemented the simplest form of the Heisenberg model, expressed as:

\begin{equation}
    H = - J \sum_{i \neq j} \vv{S_{i}} \cdot \vv{S_{j}},
    \label{eq_heinsen}
\end{equation}

where $\vv{S_{i}}$ ($\vv{S_{j}}$) represents the spin-$5/2$ vector oriented along the atomic moment at site $i(j)$, and $J$ denotes the exchange coupling parameter between the magnetic moments. 

\begin{figure*}[t]
    \centering
    \includegraphics[scale=0.85]{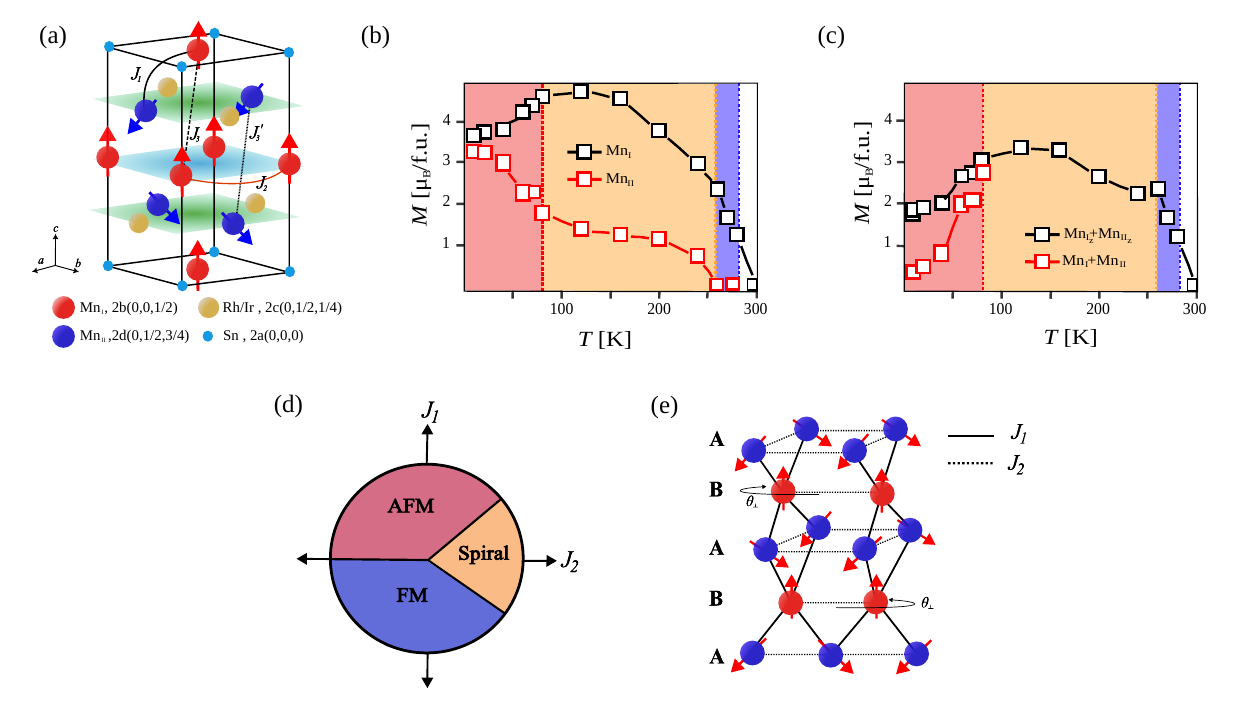}
    \caption{(a) Crystal structure of tetragonal Mn$_{2}$XSn (X = Rh, Sn) compounds, showing the spin orientations in the non-collinear state. The exchange parameters $J_i$ are illustrated, which couple the Mn moments on similar and on inequivalent sublattices. (b)-(c) Experimental reported temperature-dependent evolution of the magnetic moments in \mrs, illustrating the different magnetic phases: canted (region shaded in red), collinear ferrimagnetic (orange region), and disordered paramagnetic (blue region). (b) Temperature-driven realignment of magnetic moments. (c) Temperature evolution of the z-components of the Mn$_{I}$ and Mn$_{II}$ moments. The data are reproduced from Ref.~\cite{meshcheriakova2014large} (Mn$_{2}$RhSn compound only) using the Engauge Digitizer software ~\cite{engauge_digitzer_wojtyniak2020data}. (d) Magnetic ground states as a function of the $J_1$ and $J_2$ exchange coupling. (e) Spiral magnetic structure.
    }
    \label{figure1}
\end{figure*}

The magnetic interactions are governed by various exchange couplings, denoted as $J_i$,  which connect the different Mn moments (whether belonging to the same or to a different sublattice). These exchange interactions are categorized based on their distance and the crystal symmetry (see Fig.~\ref{figure1}(a)). The nearest-neighbor interaction $J_1$ $(d = 2.70~\text{\r{A}})$ couples Mn moments located on different Wyckoff positions (or equivalently, different crystallographic sublattices) within the tetrahedral environment. The next-nearest-neighbor interaction $J_2$ $(d = 4.37~\text{\r{A}})$ connects Mn moments located on the same Wyckoff position but in adjacent unit cells within the {\it{ab}}-plane. Finally, $J_{3}$ and $J'_3$ $(d = 4.45~\text{\r{A}})$, despite being at the same distance, couple Mn moments on the same Wyckoff position but between different sublattices, as shown in Fig.~\ref{figure1}(a). According to the spin orientation illustrated in the magnetic lattice, the nearest-neighbor interaction $(J_1)$ promotes antiparallel alignment of the spins, while the next-nearest-neighbor interaction $(J_2)$ tends to align the spins parallel to each other. However, a difference is observed between the last couplings: the $J_{3}$ keeps Mn$_{I}$ moments to be aligned parallel, while the $J'_3$ interaction between Mn$_{II}$ moments aligns them to be antiparallel to each other. This provides insight into the competing exchange coupling and the different magnetic regimes which could arise from these interactions.

\begin{figure*}[t]
    \centering
    \includegraphics[scale=0.90]{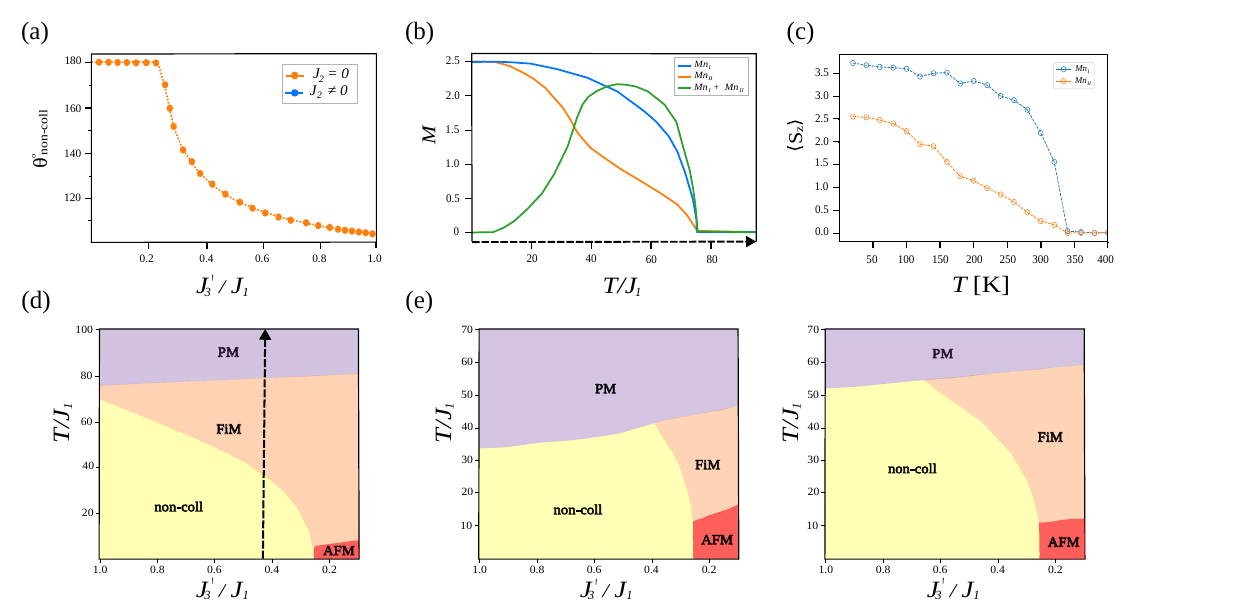}
    \caption{(a) Non-collinear angle as a function of $J'_3/J_1$. (b) Magnetic moments per site and the total magnetic moment as a function of temperature at $J'_3/J_1 = 0.424$. (c) Temperature dependence of the z-component of the spin expectation value for each magnetic site. The critical temperature is $T_\textit{C} = 350$~K. (d) Phase diagram as a function of temperature and the $J'_3/J_1$ exchange interaction, identifying four distinct magnetic phases: 1. Non-collinear (non-coll), 2. Antiferromagnetic (AFM), 3. Ferrimagnetic (FiM), and 4. Paramagnetic (PM). The dashed line indicates the cut for graph b. (e) Phase diagram as a function of temperature and $J'_3/J_1$, for $J_3 = 0.1$ (left) and $J_3 = 0.5$ (right). 
    }
    \label{figure2}
\end{figure*}

We determine the mean-field ground state of the model in Eq.~\ref{eq_heinsen} using a self-consistent approach starting from random spin orientations. We explore the role of each J-interaction utilizing  several regimes. First, we include $J_1$ and $J_2$ with the parametrization: $J_1 = \sin(\theta)$ and $J_2 = \cos(\theta)$, with $\theta$ ranging from $0^{\circ}$ to $360^{\circ}$. In Fig.~\ref{figure1}(d), we present the magnetic configurations resulting from the competition between these two couplings. In cases where $J_1$ or $J_2$ becomes significantly dominant, we observe an AFM or FM state. However, within the range $0.5 \leq J_1/J_2 \leq 2.5$, our simulations reveal the formation of a spiral (frustrated magnetic) state (see Fig.~\ref{figure1}(e)). In this spiral state, an AB stacking pattern is observed, where A corresponds to the Mn$_{II}$ sites and B to the Mn$_{I}$ sites, with each B layer rotated relative to the Mn$_{I}$ sites. In this configuration, each spin is rotated by $\theta = 120^{\circ}$ relative to its nearest neighbor, indicating a non-collinear configuration analogous to geometrically frustrated triangular lattices ~\cite{triangulo_1_balents2010spin,triangulo_2_starykh2015unusual}.



As next, we consider remaining two exchange couplings, $J_{3}$ and $J'_3$, which represent interactions between Mn moments on the same Wyckoff position but different sublattices, as previously described. To explore their impact, we start from a spiral (non-collinear) state obtained in the intermediate $J_1$/$J_2$ regime and examine how this state evolves with additional inter-sublattice couplings. Initially, we fix the value of $J'_3$ and compute the canting angle $(\theta_{non-coll})$ as a function of the $J_3$ interaction strength. Our results (see Appendix~\ref{append_1}) indicate that canting angle remains unchanged across the entire range of $J_3$ values. However, upon fixing $J_3$, we observe a variation in the canting angle as a function of  $J'_3$ (see Fig.~\ref{figure2}(a)), where several degrees of canting can be achieved as a function of the coupling between the sites. Interestingly, the $J_2$ exchange parameter does not affect the canting angle within our model, likely due to its intra-sublattice nature, which does not directly affect the inter-sublattice spin arrangement, unlike $J_{3}$ and $J'_3$.

Once the ground state of our model is established, we compute a phase diagram as a function of the temperature and $J'_3$ (see Fig.~\ref{figure2}(d)). We identified four distinct magnetic phases: (1) Non-collinear (non-coll, yellow): characterized by a canting angle between sublattices, measured along the z-axis; (2) Antiferromagnetic (AFM, red): a fully compensated unit cell where the sum of the spin directions equals zero; (3) Ferrimagnetic (FiM, orange): a unit cell where spins between sublattices are antiparallel but the net spin within the cell is non-zero; and (4) Paramagnetic (PM, purple): where no net spin magnitude or direction is present.  We have further explored the impact of $J_3$ on these magnetic states. In Fig.~\ref{figure2}(e), we tested the dependence with respect to the strength of the $J_3$ exchange. The FiM phase is suppressed when $J_3$ is low ($J_3 = 0.1$, left) compared to when $J_3 = 0.5$ (right). This insight suggests that the FiM phase emerges when the interaction between Mn$_{I}$ sites is sufficiently strong, leading to a non-compensated unit cell; on the other hand, we obtain a direct transition from the canted phase to the paramagnetic phase. All simulations were performed for a fixed value of $J_2$ different to zero, as no change in the canting angle were observed in its presence. However, in Appendix~\ref{append_2}, we present the phase diagram excluding this interaction and note that the stability of all the phases are reduced for all the phases found. These results indicate that the semi-quantitative phase diagram can be captured within a mean field theory using only isotropic couplings. No anisotropies are needed to account for the experimental findings.

According to experimental reports ~\cite{meshcheriakova2014large}, the canting angle is expected to be $\theta = 125^\circ$ (along the $z$-axis) at $T \leq 80$ K. This corresponds to $J_{3}^{'} = 0.424$ in our model and phase diagram (see dashed line in Fig.~\ref{figure2}(d)). We plotted the evolution of the spin moments as a function of the temperature for different magnetic sites, along with the total sum of these moments (see Fig.~\ref{figure2}(b)). This behavior aligns closely with the experimental findings reported by Meshcheriakova \textit{et al.} (depicted in Fig.~\ref{figure1}(b,c)), which show similar variations in magnetic moments. The consistency between our model and experimental observations confirms that our approach effectively captures the non-collinear magnetic nature expected in this inverse Heusler ferrimagnet. Specifically, when $J'_3$ exchange is significant, the system transitions to a non-collinear regime. Furthermore, as the temperature increases, a ferrimagnetic regime emerges, consistent with experimental trends.

\section{\textit{Ab initio} simulations} \label{abinitio}

To validate our ground state and investigate the electronic structure, we performed density functional theory (DFT) calculations as implemented in VASP~\cite{vasp_1kresse1996efficient,vasp_2blochl1994projector}. We used the local spin density approximation (LSDA+U) for the description of the exchange and correlation energy. The Hubbard $U$ correction was applied to the Mn-$3d$ states following the Dudarev implementation, with an effective parameter of $U_{\rm eff}=2.0$~eV~\cite{lsda_1_dudarev1998electron,lsda_2_petukhov2003correlated}. A plane-wave cutoff energy of 400~eV ensured well-converged results, and the Brillouin zone (BZ) was sampled with a $\Gamma$-centered 8x8x5 $k$-point mesh. For further analysis, we extract Wannier functions from the DFT band structure via the WANNIER90 package ~\cite{wannier90_mostofi2014updated} with initial projections onto the $d$ orbitals of Mn, Rh, and the $p$ orbitals of Sn. We then find the locations of Weyl points and present surface state calculations as implemented in WannierTools ~\cite{WannierTools_wu2018wanniertools}.

\begin{figure*}
    \centering
    \includegraphics[scale=0.80]{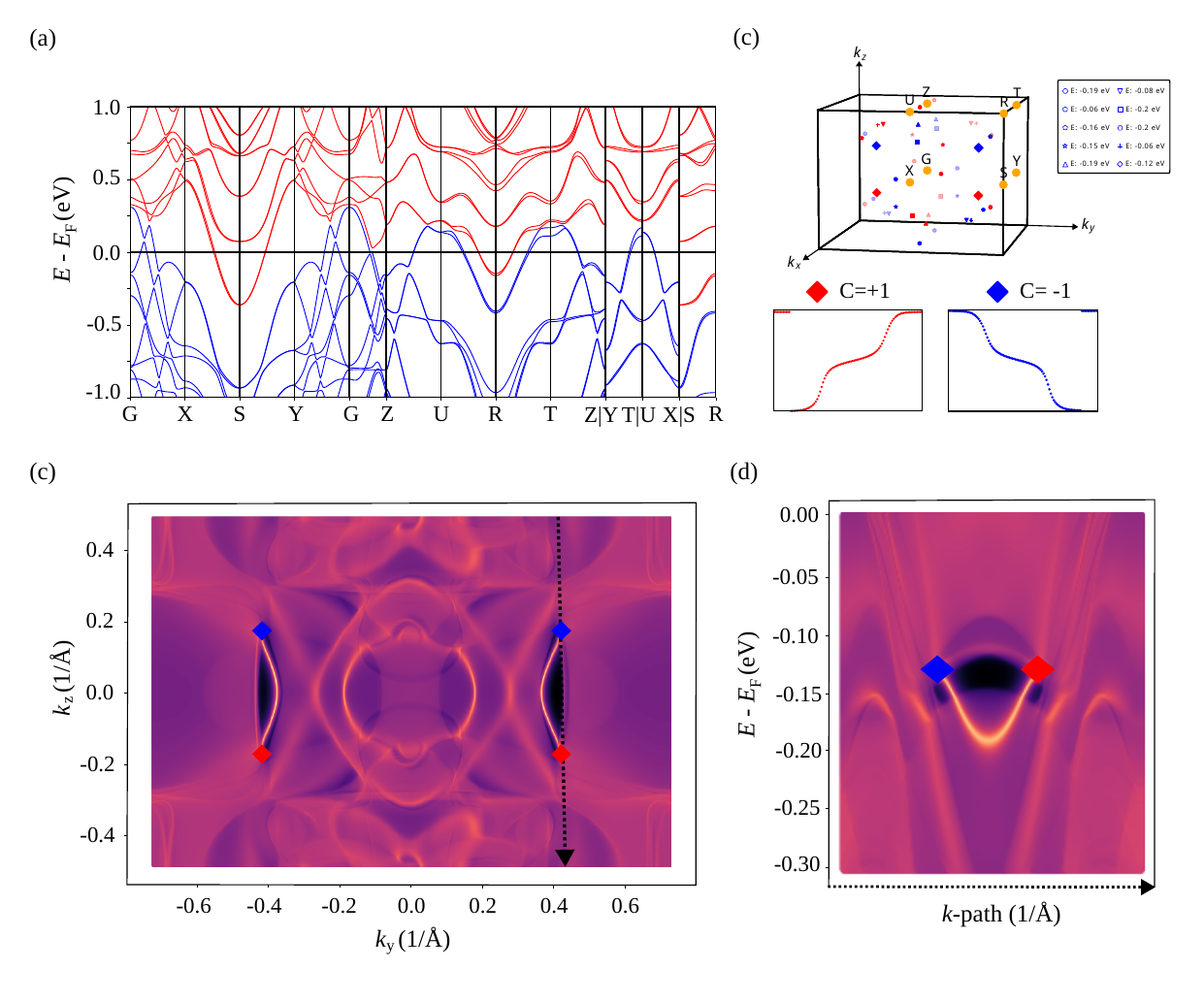}
    \caption{(a) Electronic band structure of Mn$_{2}$RhSn. The blue (red) color stands for the valence (conduction) bands according to the band index. (b) BZ and the spatial distribution of Weyl nodes. In red (blue) Weyl nodes with $+1$ ($-1$). The legend indicates the marker used for each family of nodes. (c) DFT Fermi surface projection along the [100] surface. (d) Calculated spectral function along the $k$-path in the [100] projection shown in (c), with the two Weyl points highlighted.}
    
    \label{figure4}
\end{figure*}

In addition, we evaluated Heisenberg exchange constants using the magnetic force theorem as implemented within multiple scattering theory~\cite{Liechtenstein1987,Hoffmann2020}. Using the magnetic force theorem, Mn$_2$RhSn was considered in a paramagnetic configuration, since its ground state at $T = 0$~K is non-collinear. Above $T_\text{C}$, in the paramagnetic state, the probability of finding a local moment oriented along a particular direction is the same for all orientations, and consequently the average magnetization per site is zero. Nevertheless, local magnetic moments are formed and the related paramagnetic susceptibility reflects the ground-state spin configuration. The paramagnetic state was simulated in the framework of the disordered local moment theory as it is implemented within the multiple scattering theory~\cite{Gyorffy1985,Staunton1985}.

Our \textit{ab initio} calculations yield a non-collinear ground state with out-of-plane magnetic moments, consistent with previous studies ~\cite{meshcheriakova2014large}. The calculated canting angle is $\theta = (180 \pm 35)^{\circ}$, with a total unit cell magnetization of $M \approx 1.92~\mu_\text{B}$. In agreement with our mean-field approximation at $k_\text{B}T \approx 0.1$ and $J_{3}^{'} = 0.424$, showing identical spin orientation. These findings are summarized in the Appendix~\ref{append_3}. From the magnetic force theorem, we report the exchange couplings summarized in Table \ref{table_exchange}. These values are consistent with our mean-field approximation, particularly in capturing the competing exchanges between the sublattices. In Section \ref{mean-field} and Fig.~\ref{figure1}(a), we present simulations where $J'_3$ or $J_{3}$ are fixed. However, we also explore scenarios as a competition between these two interactions (ratio), as described in the Appendix~\ref{append_4}. Our mean-field model predicts that the non-collinear angle observed in experiments is achieved when the ratio lies within $-1.5 \leq J_{3}^{'} / J_{3} \leq -2.08$, consistent with the results of the magnetic force theorem. Finally, we computed the critical temperature using the extracted exchange constants from the magnetic force theorem (see Fig.~\ref{figure2}(c)). The computed critical temperature is found to be 330~K, and a spin reorientation from a non-collinear to a collinear ferrimagnetic state occurs as the temperature increases, which aligns with our results. These findings demonstrate that, as long as interactions between Mn$_{II}$ sublattices is significant, (Mn$_2$-based) Heusler magnetic structures are expected to stabilize in a canted regime. As the ingredients of the model are general enough, we anticipate that it could be applicable to the broader family of tetragonal inverse Heusler systems ~\cite{mn_ir_sn_mozur2024magnetic,heusler_mn2pt_giri2020robust}.

\begin{table}[h]
\begin{tabular}{|c|c|c|}
\hline
\textbf{\begin{tabular}[c]{@{}l@{}}Exchange \\ Coupling (J)\end{tabular}} & \textbf{Distance (\text{\r{A}})} & \textbf{J (meV)} \\ \hline
\textbf{$J_1$}                                                             & 2.70                  & -4.7648          \\ \hline
\textbf{$J_2$}                                                             & 4.37                  & 0.5376           \\ \hline
\textbf{$J_3$}                                                             & 4.37                  & 0.5680            \\ \hline
\textbf{$J'_3$}                                                             & 4.45                  & -0.3664           \\ \hline
\end{tabular}
\caption{Exchange couplings extracted from the magnetic force theorem.}
\label{table_exchange}
\end{table}

\section{Topological analysis: Weyl nodes and surface states} \label{topological}

In Section \ref{mean-field}, we introduced the non-centrosymmetric crystal structure of Mn$_2$(Rh/Ir)Sn. The onset of magnetic order reduces the symmetry. Specifically, the four-fold axis along the $z$-direction, the two-fold axes in the $x-y$ plane, and time-reversal symmetry are all broken. The only remaining unitary symmetries are the mirror symmetry $\{m_{100} \mid 0\}$ and the glide symmetries $\{m_{010} \mid 1/2, 0, 1/2\}$, $\{2_{001} \mid 1/2, 0, 1/2\}$. These characteristics are encapsulated in the magnetic space group $Pm'n'2_{1}$ (No. 31.127). After identifying the symmetry and space group, we perform two non-self-consistent calculations: one along the high-symmetry lines to obtain the electronic band dispersion (see Fig.~\ref{figure4}(a)), and another at the high-symmetry points to calculate the wavefunctions. For the second calculation, we use \Verb|IrRep|  to extract the magnetic corepresentations and the band representations (BR) of the occupied bands, which are then uploaded to the BCS server for symmetry-based topological predictions ~\cite{mvasp2trac1_xu2020high,mvasp2trac2_elcoro2021magnetic,iraola2022irrep}. \mrs\ is classified as a trivial insulator with symmetry indicators at $\delta_{(1)}^{a}=1$.

By analyzing our Brillouin zone (BZ), we identify Weyl nodes near the Fermi level, grouped into families according to their energies, with their locations shown in Fig.~\ref{figure4}(b). Each family consists of four nodes related by symmetry: two with a topological charge of $+1$ and two with $-1$, resulting in a net topological charge of zero for each family. We further confirm their topological charge by calculating Wilson loops on a sphere surrounding the degeneracy points. Our results reveal that pairs of Weyl points with opposite charges lie along a line, potentially giving rise to topological Fermi arcs across the BZ.

We compute the [100] cleave plane and project the Weyl points onto this surface, as illustrated in Fig.~\ref{figure4}(c). Our simulations show that the families of Weyl points at $E_{W} = -0.12$ eV coincide with the surface state calculations, where regions of very high density connect these Weyl points along $k_{z}$. In Fig.~\ref{figure4}(d), we display the (100) surface spectrum along the path shown in Fig.~\ref{figure4}(c). This path crosses both Weyl points with opposite topological charges, revealing a bright surface state that connects the projections of these nodes. This observation supports the existence of Fermi arcs originating from these projections.

\section{Conclusions} \label{conclusions}

In this work, we have investigated the non-collinear spin configurations found in Mn$_2$-based Heusler magnets. Using a mean-field approximation, we demonstrated how competing interactions between the out-of-plane sublattices lead to the canted angle observed in experiments. Additionally, we computed and analyzed the exchange couplings using \textit{ab initio} methods, confirming that our approximations align with established methodologies. We explored various magnetic regimes, providing general insights that can be extended to other Mn$_2$-based compounds. Finally, we identified the Weyl points and Fermi arcs emerging from the [100] surface in Mn$_2$RhSn, highlighting the topological properties inherent to these systems. The presence of these topological properties, along with the varying degrees of spin canting in inverse Heusler magnets, could give rise to distinct quantum transport phenomena, which provides insight for future research directions.

\section{Acknowledgements} 
M.G.V. thanks support to the Spanish Ministerio de Ciencia e Innovaci\'on (PID2022-142008NB-I00), the Canada Excellence Research Chairs Program for Topological Quantum Matter,  the IKUR Strategy under the collaboration agreement between Ikerbasque Foundation and DIPC on behalf of the Department of Education of the Basque Government and to Diputaci\'on Foral de Gipuzkoa Programa Mujeres y Ciencia. We acknowledge financial support by the Deutsche Forschungsgemeinschaft (DFG, German Research Foundation), through the Würzburg-Dresden Cluster of Excellence on Complexity and Topology in Quantum Matter, ct.qmat (EXC 2147, Project ID 390858490) and the GA 3314/1-1 – FOR 5249 (QUAST). A.E. acknowledges funding from the Fonds zur Förderung der Wissenschaftlichen Forschung (FWF) under Grant No. I 5384. We acknowledge Anastasios Markou, Iñigo Robredo and Rebeca Ibarra for discussions.
\appendix

\section{} \label{append_1}

In this case, we fixed the exchange parameter $J'_3$ and computed the non-collinear angle as a function of the $J_{3}$ coupling. Our simulations indicate that the non-collinear angle remains unchanged across the entire range of $J_{3}$.

\begin{figure}[h]
    \centering
    \includegraphics{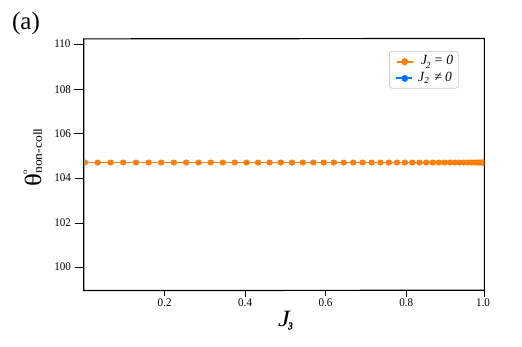}
    \caption{(a) Non-collinear angle as a function of $J_{3}$  }
    \label{S1_figure}
\end{figure}

\section{} \label{append_3}

Computed atomic magnetic moments from DFT and mean-field simulations reveal consistent magnetic moment orientations in both approaches. In the mean-field solution, a magnetic y-component emerges due to the initial random spin orientation, whereas in DFT, we impose an initial non-collinear magnetic state. Nevertheless, both methods successfully capture the underlying magnetic nature of the system.

\begin{table}[h]
\centering
\begin{tabular}{|c|ccc|ccc|}
\hline
\multirow{2}{*}{} & \multicolumn{3}{c|}{\textbf{DFT}} & \multicolumn{3}{c|}{\textbf{Mean-field}} \\ \cline{2-7} 
                  & \multicolumn{1}{c|}{x}      & \multicolumn{1}{c|}{y}    & z      & \multicolumn{1}{c|}{x}     & \multicolumn{1}{c|}{y}      & z      \\ \hline
$Mn_{I}$         & \multicolumn{1}{c|}{0.00}   & \multicolumn{1}{c|}{0.00} & 3.84   & \multicolumn{1}{c|}{0.00}  & \multicolumn{1}{c|}{0.00}   & 2.5    \\ \hline
$Mn_{I}$         & \multicolumn{1}{c|}{0.00}   & \multicolumn{1}{c|}{0.00} & 3.84   & \multicolumn{1}{c|}{0.00}  & \multicolumn{1}{c|}{0.00}   & 2.5    \\ \hline
$Mn_{II}$        & \multicolumn{1}{c|}{2.03}   & \multicolumn{1}{c|}{0.00} & -3.01  & \multicolumn{1}{c|}{-1.60} & \multicolumn{1}{c|}{1.309}  & -1.616 \\ \hline
$Mn_{II}$        & \multicolumn{1}{c|}{-2.03}  & \multicolumn{1}{c|}{0.00} & -3.01  & \multicolumn{1}{c|}{1.60}  & \multicolumn{1}{c|}{-1.309} & -1.616 \\ \hline
\end{tabular}
\label{table1}
\caption{Magnetic moments of Mn sites calculated using DFT and the Mean-Field Approximation. }
\end{table}

\section{} \label{append_2}

We computed the phase diagram as a function of temperature and $J'_3$ in the absence of $J_{2}$ coupling. Our simulations indicate that the non-collinear angle persists regardless of whether $J_{2}$ is present (see Fig. \ref{figure2} (a)). However, the phase diagram reveals that while the magnetic regime remains unchanged, the system's robustness decreases. Although the inclusion of $J_{2}$  is not crucial for a qualitative mean-field description, it may play an important role in extrapolating macroscopic properties of the system.

\begin{figure}[H]
    \centering
    \includegraphics[scale=0.8]{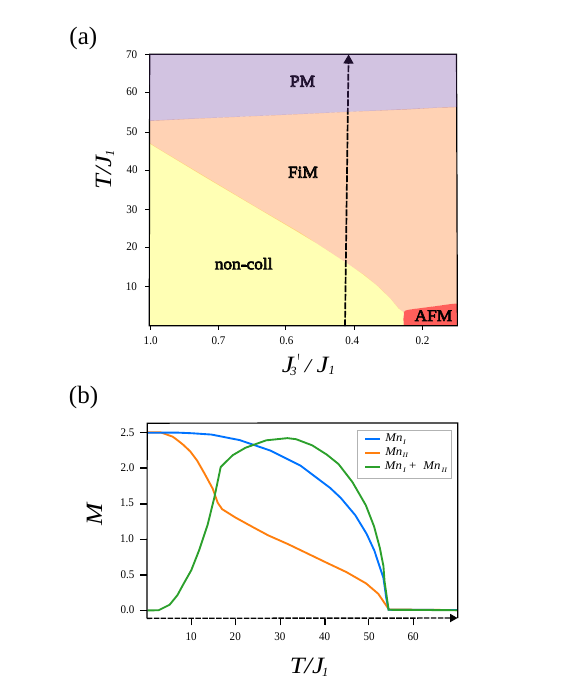}
    \caption{(a) Phase diagram as a function of temperature and the $J'_3/J_1$ exchange interaction for $J_{2} = 0$. (b) Magnetic moments per site and the total magnetic moment for $J_{2} = 0$. }
    \label{S2_figure}
\end{figure}

\section{} \label{append_4}

We computed the non-collinear angle as a function of the competition between $J_{3}$ and $J'_3$. Varying their ratio leads to different degrees of canting. This allows us to extrapolate a range of non-collinear angles depending on the exchange ratio.

\begin{figure}[h]
    \centering
    \includegraphics[scale=0.8]{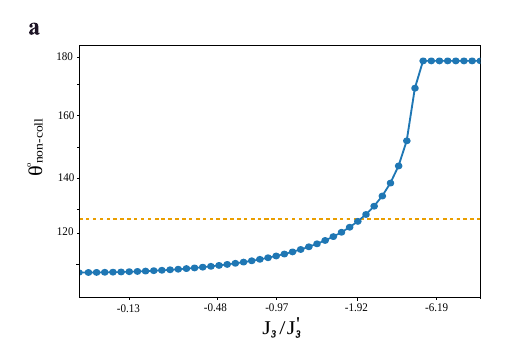}
    \caption{(a) Non-collinear angle as a function of the ratio between $J_{3}$ and $J'_3$. Dashed orange line is the experimental reported angle: $\theta = 125^{\circ}$  }
    \label{S4_figure}
\end{figure}

\bibliography{acs-achemso.bib}

\begin{thebibliography}{35}%
\makeatletter
\providecommand \@ifxundefined [1]{%
 \@ifx{#1\undefined}
}%
\providecommand \@ifnum [1]{%
 \ifnum #1\expandafter \@firstoftwo
 \else \expandafter \@secondoftwo
 \fi
}%
\providecommand \@ifx [1]{%
 \ifx #1\expandafter \@firstoftwo
 \else \expandafter \@secondoftwo
 \fi
}%
\providecommand \natexlab [1]{#1}%
\providecommand \enquote  [1]{``#1''}%
\providecommand \bibnamefont  [1]{#1}%
\providecommand \bibfnamefont [1]{#1}%
\providecommand \citenamefont [1]{#1}%
\providecommand \href@noop [0]{\@secondoftwo}%
\providecommand \href [0]{\begingroup \@sanitize@url \@href}%
\providecommand \@href[1]{\@@startlink{#1}\@@href}%
\providecommand \@@href[1]{\endgroup#1\@@endlink}%
\providecommand \@sanitize@url [0]{\catcode `\\12\catcode `\$12\catcode `\&12\catcode `\#12\catcode `\^12\catcode `\_12\catcode `\%12\relax}%
\providecommand \@@startlink[1]{}%
\providecommand \@@endlink[0]{}%
\providecommand \url  [0]{\begingroup\@sanitize@url \@url }%
\providecommand \@url [1]{\endgroup\@href {#1}{\urlprefix }}%
\providecommand \urlprefix  [0]{URL }%
\providecommand \Eprint [0]{\href }%
\providecommand \doibase [0]{https://doi.org/}%
\providecommand \selectlanguage [0]{\@gobble}%
\providecommand \bibinfo  [0]{\@secondoftwo}%
\providecommand \bibfield  [0]{\@secondoftwo}%
\providecommand \translation [1]{[#1]}%
\providecommand \BibitemOpen [0]{}%
\providecommand \bibitemStop [0]{}%
\providecommand \bibitemNoStop [0]{.\EOS\space}%
\providecommand \EOS [0]{\spacefactor3000\relax}%
\providecommand \BibitemShut  [1]{\csname bibitem#1\endcsname}%
\let\auto@bib@innerbib\@empty
\bibitem [{\citenamefont {Hirohata}\ \emph {et~al.}(2020)\citenamefont {Hirohata}, \citenamefont {Yamada}, \citenamefont {Nakatani}, \citenamefont {Prejbeanu}, \citenamefont {Di{\'e}ny}, \citenamefont {Pirro},\ and\ \citenamefont {Hillebrands}}]{intro_hirohata2020review}%
  \BibitemOpen
  \bibfield  {author} {\bibinfo {author} {\bibfnamefont {A.}~\bibnamefont {Hirohata}}, \bibinfo {author} {\bibfnamefont {K.}~\bibnamefont {Yamada}}, \bibinfo {author} {\bibfnamefont {Y.}~\bibnamefont {Nakatani}}, \bibinfo {author} {\bibfnamefont {I.-L.}\ \bibnamefont {Prejbeanu}}, \bibinfo {author} {\bibfnamefont {B.}~\bibnamefont {Di{\'e}ny}}, \bibinfo {author} {\bibfnamefont {P.}~\bibnamefont {Pirro}},\ and\ \bibinfo {author} {\bibfnamefont {B.}~\bibnamefont {Hillebrands}},\ }\bibfield  {title} {\bibinfo {title} {Review on spintronics: Principles and device applications},\ }\href@noop {} {\bibfield  {journal} {\bibinfo  {journal} {Journal of Magnetism and Magnetic Materials}\ }\textbf {\bibinfo {volume} {509}},\ \bibinfo {pages} {166711} (\bibinfo {year} {2020})}\BibitemShut {NoStop}%
\bibitem [{\citenamefont {Fert}\ \emph {et~al.}(2017)\citenamefont {Fert}, \citenamefont {Reyren},\ and\ \citenamefont {Cros}}]{intro_fert2017magnetic}%
  \BibitemOpen
  \bibfield  {author} {\bibinfo {author} {\bibfnamefont {A.}~\bibnamefont {Fert}}, \bibinfo {author} {\bibfnamefont {N.}~\bibnamefont {Reyren}},\ and\ \bibinfo {author} {\bibfnamefont {V.}~\bibnamefont {Cros}},\ }\bibfield  {title} {\bibinfo {title} {Magnetic skyrmions: advances in physics and potential applications},\ }\href@noop {} {\bibfield  {journal} {\bibinfo  {journal} {Nature Reviews Materials}\ }\textbf {\bibinfo {volume} {2}},\ \bibinfo {pages} {1} (\bibinfo {year} {2017})}\BibitemShut {NoStop}%
\bibitem [{\citenamefont {Tokura}\ and\ \citenamefont {Kanazawa}(2020)}]{intro_tokura2020magnetic}%
  \BibitemOpen
  \bibfield  {author} {\bibinfo {author} {\bibfnamefont {Y.}~\bibnamefont {Tokura}}\ and\ \bibinfo {author} {\bibfnamefont {N.}~\bibnamefont {Kanazawa}},\ }\bibfield  {title} {\bibinfo {title} {Magnetic skyrmion materials},\ }\href@noop {} {\bibfield  {journal} {\bibinfo  {journal} {Chemical Reviews}\ }\textbf {\bibinfo {volume} {121}},\ \bibinfo {pages} {2857} (\bibinfo {year} {2020})}\BibitemShut {NoStop}%
\bibitem [{\citenamefont {Graf}\ \emph {et~al.}(2011)\citenamefont {Graf}, \citenamefont {Felser},\ and\ \citenamefont {Parkin}}]{heusler1_graf2011simple}%
  \BibitemOpen
  \bibfield  {author} {\bibinfo {author} {\bibfnamefont {T.}~\bibnamefont {Graf}}, \bibinfo {author} {\bibfnamefont {C.}~\bibnamefont {Felser}},\ and\ \bibinfo {author} {\bibfnamefont {S.~S.}\ \bibnamefont {Parkin}},\ }\bibfield  {title} {\bibinfo {title} {Simple rules for the understanding of heusler compounds},\ }\href@noop {} {\bibfield  {journal} {\bibinfo  {journal} {Progress in solid state chemistry}\ }\textbf {\bibinfo {volume} {39}},\ \bibinfo {pages} {1} (\bibinfo {year} {2011})}\BibitemShut {NoStop}%
\bibitem [{\citenamefont {Wollmann}\ \emph {et~al.}(2017)\citenamefont {Wollmann}, \citenamefont {Nayak}, \citenamefont {Parkin},\ and\ \citenamefont {Felser}}]{heusler2_wollmann2017heusler}%
  \BibitemOpen
  \bibfield  {author} {\bibinfo {author} {\bibfnamefont {L.}~\bibnamefont {Wollmann}}, \bibinfo {author} {\bibfnamefont {A.~K.}\ \bibnamefont {Nayak}}, \bibinfo {author} {\bibfnamefont {S.~S.}\ \bibnamefont {Parkin}},\ and\ \bibinfo {author} {\bibfnamefont {C.}~\bibnamefont {Felser}},\ }\bibfield  {title} {\bibinfo {title} {Heusler 4.0: tunable materials},\ }\href@noop {} {\bibfield  {journal} {\bibinfo  {journal} {Annual Review of Materials Research}\ }\textbf {\bibinfo {volume} {47}},\ \bibinfo {pages} {247} (\bibinfo {year} {2017})}\BibitemShut {NoStop}%
\bibitem [{\citenamefont {Palmstr{\o}m}(2016)}]{heusler_5_palmstrom2016heusler}%
  \BibitemOpen
  \bibfield  {author} {\bibinfo {author} {\bibfnamefont {C.~J.}\ \bibnamefont {Palmstr{\o}m}},\ }\bibfield  {title} {\bibinfo {title} {Heusler compounds and spintronics},\ }\href@noop {} {\bibfield  {journal} {\bibinfo  {journal} {Progress in Crystal Growth and Characterization of Materials}\ }\textbf {\bibinfo {volume} {62}},\ \bibinfo {pages} {371} (\bibinfo {year} {2016})}\BibitemShut {NoStop}%
\bibitem [{\citenamefont {Picozzi}\ \emph {et~al.}(2004)\citenamefont {Picozzi}, \citenamefont {Continenza},\ and\ \citenamefont {Freeman}}]{heusler_3_co2MnSipicozzi2004role}%
  \BibitemOpen
  \bibfield  {author} {\bibinfo {author} {\bibfnamefont {S.}~\bibnamefont {Picozzi}}, \bibinfo {author} {\bibfnamefont {A.}~\bibnamefont {Continenza}},\ and\ \bibinfo {author} {\bibfnamefont {A.}~\bibnamefont {Freeman}},\ }\bibfield  {title} {\bibinfo {title} {Role of structural defects on the half-metallic character of {Co$_{2}$MnGe} and {Co$_{2}$MnSi} heusler alloys},\ }\href@noop {} {\bibfield  {journal} {\bibinfo  {journal} {Physical Review B}\ }\textbf {\bibinfo {volume} {69}},\ \bibinfo {pages} {094423} (\bibinfo {year} {2004})}\BibitemShut {NoStop}%
\bibitem [{\citenamefont {Ludbrook}\ \emph {et~al.}(2017)\citenamefont {Ludbrook}, \citenamefont {Ruck},\ and\ \citenamefont {Granville}}]{heusler_7_co2mnga_ludbrook2017perpendicular}%
  \BibitemOpen
  \bibfield  {author} {\bibinfo {author} {\bibfnamefont {B.}~\bibnamefont {Ludbrook}}, \bibinfo {author} {\bibfnamefont {B.}~\bibnamefont {Ruck}},\ and\ \bibinfo {author} {\bibfnamefont {S.}~\bibnamefont {Granville}},\ }\bibfield  {title} {\bibinfo {title} {Perpendicular magnetic anisotropy in {Co$_{2}$MnGa} and its anomalous hall effect},\ }\href@noop {} {\bibfield  {journal} {\bibinfo  {journal} {Applied Physics Letters}\ }\textbf {\bibinfo {volume} {110}} (\bibinfo {year} {2017})}\BibitemShut {NoStop}%
\bibitem [{\citenamefont {Guin}\ \emph {et~al.}(2019)\citenamefont {Guin}, \citenamefont {Vir}, \citenamefont {Zhang}, \citenamefont {Kumar}, \citenamefont {Watzman}, \citenamefont {Fu}, \citenamefont {Liu}, \citenamefont {Manna}, \citenamefont {Schnelle}, \citenamefont {Gooth} \emph {et~al.}}]{weyl_guin2019zero}%
  \BibitemOpen
  \bibfield  {author} {\bibinfo {author} {\bibfnamefont {S.~N.}\ \bibnamefont {Guin}}, \bibinfo {author} {\bibfnamefont {P.}~\bibnamefont {Vir}}, \bibinfo {author} {\bibfnamefont {Y.}~\bibnamefont {Zhang}}, \bibinfo {author} {\bibfnamefont {N.}~\bibnamefont {Kumar}}, \bibinfo {author} {\bibfnamefont {S.~J.}\ \bibnamefont {Watzman}}, \bibinfo {author} {\bibfnamefont {C.}~\bibnamefont {Fu}}, \bibinfo {author} {\bibfnamefont {E.}~\bibnamefont {Liu}}, \bibinfo {author} {\bibfnamefont {K.}~\bibnamefont {Manna}}, \bibinfo {author} {\bibfnamefont {W.}~\bibnamefont {Schnelle}}, \bibinfo {author} {\bibfnamefont {J.}~\bibnamefont {Gooth}}, \emph {et~al.},\ }\bibfield  {title} {\bibinfo {title} {Zero-field nernst effect in a ferromagnetic kagome-lattice weyl-semimetal {Co$_{3}$Sn$_{2}$S$_{2}$}},\ }\href@noop {} {\bibfield  {journal} {\bibinfo  {journal} {Advanced Materials}\ }\textbf {\bibinfo {volume} {31}},\ \bibinfo {pages} {1806622} (\bibinfo {year} {2019})}\BibitemShut {NoStop}%
\bibitem [{\citenamefont {K{\"u}bler}\ and\ \citenamefont {Felser}(2018)}]{heusler_4_mn3Sn}%
  \BibitemOpen
  \bibfield  {author} {\bibinfo {author} {\bibfnamefont {J.}~\bibnamefont {K{\"u}bler}}\ and\ \bibinfo {author} {\bibfnamefont {C.}~\bibnamefont {Felser}},\ }\bibfield  {title} {\bibinfo {title} {Weyl fermions in antiferromagnetic {Mn$_{3}$Sn} and {Mn$_{3}$Ge}},\ }\href@noop {} {\bibfield  {journal} {\bibinfo  {journal} {Europhysics Letters}\ }\textbf {\bibinfo {volume} {120}},\ \bibinfo {pages} {47002} (\bibinfo {year} {2018})}\BibitemShut {NoStop}%
\bibitem [{\citenamefont {K{\"u}bler}\ and\ \citenamefont {Felser}(2014)}]{hall_kubler2014non}%
  \BibitemOpen
  \bibfield  {author} {\bibinfo {author} {\bibfnamefont {J.}~\bibnamefont {K{\"u}bler}}\ and\ \bibinfo {author} {\bibfnamefont {C.}~\bibnamefont {Felser}},\ }\bibfield  {title} {\bibinfo {title} {Non-collinear antiferromagnets and the anomalous hall effect},\ }\href@noop {} {\bibfield  {journal} {\bibinfo  {journal} {Europhysics Letters}\ }\textbf {\bibinfo {volume} {108}},\ \bibinfo {pages} {67001} (\bibinfo {year} {2014})}\BibitemShut {NoStop}%
\bibitem [{\citenamefont {Felser}\ \emph {et~al.}(2015)\citenamefont {Felser}, \citenamefont {Wollmann}, \citenamefont {Chadov}, \citenamefont {Fecher},\ and\ \citenamefont {Parkin}}]{felser2015basics}%
  \BibitemOpen
  \bibfield  {author} {\bibinfo {author} {\bibfnamefont {C.}~\bibnamefont {Felser}}, \bibinfo {author} {\bibfnamefont {L.}~\bibnamefont {Wollmann}}, \bibinfo {author} {\bibfnamefont {S.}~\bibnamefont {Chadov}}, \bibinfo {author} {\bibfnamefont {G.~H.}\ \bibnamefont {Fecher}},\ and\ \bibinfo {author} {\bibfnamefont {S.~S.}\ \bibnamefont {Parkin}},\ }\bibfield  {title} {\bibinfo {title} {Basics and prospective of magnetic heusler compounds},\ }\href@noop {} {\bibfield  {journal} {\bibinfo  {journal} {APL materials}\ }\textbf {\bibinfo {volume} {3}} (\bibinfo {year} {2015})}\BibitemShut {NoStop}%
\bibitem [{\citenamefont {Manna}\ \emph {et~al.}(2018)\citenamefont {Manna}, \citenamefont {Sun}, \citenamefont {Muechler}, \citenamefont {K{\"u}bler},\ and\ \citenamefont {Felser}}]{manna2018heusler}%
  \BibitemOpen
  \bibfield  {author} {\bibinfo {author} {\bibfnamefont {K.}~\bibnamefont {Manna}}, \bibinfo {author} {\bibfnamefont {Y.}~\bibnamefont {Sun}}, \bibinfo {author} {\bibfnamefont {L.}~\bibnamefont {Muechler}}, \bibinfo {author} {\bibfnamefont {J.}~\bibnamefont {K{\"u}bler}},\ and\ \bibinfo {author} {\bibfnamefont {C.}~\bibnamefont {Felser}},\ }\bibfield  {title} {\bibinfo {title} {Heusler, weyl and berry},\ }\href@noop {} {\bibfield  {journal} {\bibinfo  {journal} {Nature Reviews Materials}\ }\textbf {\bibinfo {volume} {3}},\ \bibinfo {pages} {244} (\bibinfo {year} {2018})}\BibitemShut {NoStop}%
\bibitem [{\citenamefont {Meshcheriakova}\ \emph {et~al.}(2014)\citenamefont {Meshcheriakova}, \citenamefont {Chadov}, \citenamefont {Nayak}, \citenamefont {R{\"o}{\ss}ler}, \citenamefont {K{\"u}bler}, \citenamefont {Andr{\'e}}, \citenamefont {Tsirlin}, \citenamefont {Kiss}, \citenamefont {Hausdorf}, \citenamefont {Kalache} \emph {et~al.}}]{meshcheriakova2014large}%
  \BibitemOpen
  \bibfield  {author} {\bibinfo {author} {\bibfnamefont {O.}~\bibnamefont {Meshcheriakova}}, \bibinfo {author} {\bibfnamefont {S.}~\bibnamefont {Chadov}}, \bibinfo {author} {\bibfnamefont {A.}~\bibnamefont {Nayak}}, \bibinfo {author} {\bibfnamefont {U.}~\bibnamefont {R{\"o}{\ss}ler}}, \bibinfo {author} {\bibfnamefont {J.}~\bibnamefont {K{\"u}bler}}, \bibinfo {author} {\bibfnamefont {G.}~\bibnamefont {Andr{\'e}}}, \bibinfo {author} {\bibfnamefont {A.}~\bibnamefont {Tsirlin}}, \bibinfo {author} {\bibfnamefont {J.}~\bibnamefont {Kiss}}, \bibinfo {author} {\bibfnamefont {S.}~\bibnamefont {Hausdorf}}, \bibinfo {author} {\bibfnamefont {A.}~\bibnamefont {Kalache}}, \emph {et~al.},\ }\bibfield  {title} {\bibinfo {title} {Large noncollinearity and spin reorientation in the novel {Mn$_{2}$RhSn} heusler magnet},\ }\href@noop {} {\bibfield  {journal} {\bibinfo  {journal} {Physical review letters}\ }\textbf {\bibinfo {volume} {113}},\ \bibinfo {pages} {087203} (\bibinfo {year} {2014})}\BibitemShut {NoStop}%
\bibitem [{\citenamefont {Mozur}\ and\ \citenamefont {Seshadri}(2024)}]{mn_ir_sn_mozur2024magnetic}%
  \BibitemOpen
  \bibfield  {author} {\bibinfo {author} {\bibfnamefont {E.~M.}\ \bibnamefont {Mozur}}\ and\ \bibinfo {author} {\bibfnamefont {R.}~\bibnamefont {Seshadri}},\ }\bibfield  {title} {\bibinfo {title} {Magnetic tunability in tetragonal {Mn-Rh-Ir-Sn} inverse heusler compounds},\ }\href@noop {} {\bibfield  {journal} {\bibinfo  {journal} {Journal of Physics: Condensed Matter}\ }\textbf {\bibinfo {volume} {36}},\ \bibinfo {pages} {195802} (\bibinfo {year} {2024})}\BibitemShut {NoStop}%
\bibitem [{\citenamefont {Meinert}\ \emph {et~al.}(2011)\citenamefont {Meinert}, \citenamefont {Schmalhorst},\ and\ \citenamefont {Reiss}}]{fim_col_meinert2011exchange}%
  \BibitemOpen
  \bibfield  {author} {\bibinfo {author} {\bibfnamefont {M.}~\bibnamefont {Meinert}}, \bibinfo {author} {\bibfnamefont {J.-M.}\ \bibnamefont {Schmalhorst}},\ and\ \bibinfo {author} {\bibfnamefont {G.}~\bibnamefont {Reiss}},\ }\bibfield  {title} {\bibinfo {title} {Exchange interactions and curie temperatures of {Mn$_{2}$CoZ} compounds},\ }\href@noop {} {\bibfield  {journal} {\bibinfo  {journal} {Journal of Physics: Condensed Matter}\ }\textbf {\bibinfo {volume} {23}},\ \bibinfo {pages} {116005} (\bibinfo {year} {2011})}\BibitemShut {NoStop}%
\bibitem [{\citenamefont {Wollmann}\ \emph {et~al.}(2015)\citenamefont {Wollmann}, \citenamefont {Chadov}, \citenamefont {K{\"u}bler},\ and\ \citenamefont {Felser}}]{wollmann2015magnetism}%
  \BibitemOpen
  \bibfield  {author} {\bibinfo {author} {\bibfnamefont {L.}~\bibnamefont {Wollmann}}, \bibinfo {author} {\bibfnamefont {S.}~\bibnamefont {Chadov}}, \bibinfo {author} {\bibfnamefont {J.}~\bibnamefont {K{\"u}bler}},\ and\ \bibinfo {author} {\bibfnamefont {C.}~\bibnamefont {Felser}},\ }\bibfield  {title} {\bibinfo {title} {Magnetism in tetragonal manganese rich heusler compounds},\ }\href@noop {} {\bibfield  {journal} {\bibinfo  {journal} {Physical Review B}\ }\textbf {\bibinfo {volume} {92}},\ \bibinfo {pages} {064417} (\bibinfo {year} {2015})}\BibitemShut {NoStop}%
\bibitem [{\citenamefont {Faleev}\ \emph {et~al.}(2017)\citenamefont {Faleev}, \citenamefont {Ferrante}, \citenamefont {Jeong}, \citenamefont {Samant}, \citenamefont {Jones},\ and\ \citenamefont {Parkin}}]{faleev2017origin}%
  \BibitemOpen
  \bibfield  {author} {\bibinfo {author} {\bibfnamefont {S.~V.}\ \bibnamefont {Faleev}}, \bibinfo {author} {\bibfnamefont {Y.}~\bibnamefont {Ferrante}}, \bibinfo {author} {\bibfnamefont {J.}~\bibnamefont {Jeong}}, \bibinfo {author} {\bibfnamefont {M.~G.}\ \bibnamefont {Samant}}, \bibinfo {author} {\bibfnamefont {B.}~\bibnamefont {Jones}},\ and\ \bibinfo {author} {\bibfnamefont {S.~S.}\ \bibnamefont {Parkin}},\ }\bibfield  {title} {\bibinfo {title} {Origin of the tetragonal ground state of heusler compounds},\ }\href@noop {} {\bibfield  {journal} {\bibinfo  {journal} {Physical Review Applied}\ }\textbf {\bibinfo {volume} {7}},\ \bibinfo {pages} {034022} (\bibinfo {year} {2017})}\BibitemShut {NoStop}%
\bibitem [{\citenamefont {Wojtyniak}\ \emph {et~al.}(2020)\citenamefont {Wojtyniak}, \citenamefont {Britz}, \citenamefont {Selzer}, \citenamefont {Schwab},\ and\ \citenamefont {Lehr}}]{engauge_digitzer_wojtyniak2020data}%
  \BibitemOpen
  \bibfield  {author} {\bibinfo {author} {\bibfnamefont {J.-G.}\ \bibnamefont {Wojtyniak}}, \bibinfo {author} {\bibfnamefont {H.}~\bibnamefont {Britz}}, \bibinfo {author} {\bibfnamefont {D.}~\bibnamefont {Selzer}}, \bibinfo {author} {\bibfnamefont {M.}~\bibnamefont {Schwab}},\ and\ \bibinfo {author} {\bibfnamefont {T.}~\bibnamefont {Lehr}},\ }\bibfield  {title} {\bibinfo {title} {Data digitizing: accurate and precise data extraction for quantitative systems pharmacology and physiologically-based pharmacokinetic modeling},\ }\href@noop {} {\bibfield  {journal} {\bibinfo  {journal} {CPT: Pharmacometrics \& Systems Pharmacology}\ }\textbf {\bibinfo {volume} {9}},\ \bibinfo {pages} {322} (\bibinfo {year} {2020})}\BibitemShut {NoStop}%
\bibitem [{\citenamefont {Balents}(2010)}]{triangulo_1_balents2010spin}%
  \BibitemOpen
  \bibfield  {author} {\bibinfo {author} {\bibfnamefont {L.}~\bibnamefont {Balents}},\ }\bibfield  {title} {\bibinfo {title} {Spin liquids in frustrated magnets},\ }\href@noop {} {\bibfield  {journal} {\bibinfo  {journal} {nature}\ }\textbf {\bibinfo {volume} {464}},\ \bibinfo {pages} {199} (\bibinfo {year} {2010})}\BibitemShut {NoStop}%
\bibitem [{\citenamefont {Starykh}(2015)}]{triangulo_2_starykh2015unusual}%
  \BibitemOpen
  \bibfield  {author} {\bibinfo {author} {\bibfnamefont {O.~A.}\ \bibnamefont {Starykh}},\ }\bibfield  {title} {\bibinfo {title} {Unusual ordered phases of highly frustrated magnets: a review},\ }\href@noop {} {\bibfield  {journal} {\bibinfo  {journal} {Reports on Progress in Physics}\ }\textbf {\bibinfo {volume} {78}},\ \bibinfo {pages} {052502} (\bibinfo {year} {2015})}\BibitemShut {NoStop}%
\bibitem [{\citenamefont {Kresse}\ and\ \citenamefont {Furthm{\"u}ller}(1996{\natexlab{a}})}]{vasp_1kresse1996efficient}%
  \BibitemOpen
  \bibfield  {author} {\bibinfo {author} {\bibfnamefont {G.}~\bibnamefont {Kresse}}\ and\ \bibinfo {author} {\bibfnamefont {J.}~\bibnamefont {Furthm{\"u}ller}},\ }\bibfield  {title} {\bibinfo {title} {Efficient iterative schemes for ab initio total-energy calculations using a plane-wave basis set},\ }\href@noop {} {\bibfield  {journal} {\bibinfo  {journal} {Physical Review B}\ }\textbf {\bibinfo {volume} {54}},\ \bibinfo {pages} {11169} (\bibinfo {year} {1996}{\natexlab{a}})}\BibitemShut {NoStop}%
\bibitem [{\citenamefont {Kresse}\ and\ \citenamefont {Furthm{\"u}ller}(1996{\natexlab{b}})}]{vasp_2blochl1994projector}%
  \BibitemOpen
  \bibfield  {author} {\bibinfo {author} {\bibfnamefont {G.}~\bibnamefont {Kresse}}\ and\ \bibinfo {author} {\bibfnamefont {J.}~\bibnamefont {Furthm{\"u}ller}},\ }\bibfield  {title} {\bibinfo {title} {Efficient iterative schemes for ab initio total-energy calculations using a plane-wave basis set},\ }\href@noop {} {\bibfield  {journal} {\bibinfo  {journal} {Physical Review B}\ }\textbf {\bibinfo {volume} {54}},\ \bibinfo {pages} {11169} (\bibinfo {year} {1996}{\natexlab{b}})}\BibitemShut {NoStop}%
\bibitem [{\citenamefont {Dudarev}\ \emph {et~al.}(1998)\citenamefont {Dudarev}, \citenamefont {Botton}, \citenamefont {Savrasov}, \citenamefont {Humphreys},\ and\ \citenamefont {Sutton}}]{lsda_1_dudarev1998electron}%
  \BibitemOpen
  \bibfield  {author} {\bibinfo {author} {\bibfnamefont {S.~L.}\ \bibnamefont {Dudarev}}, \bibinfo {author} {\bibfnamefont {G.~A.}\ \bibnamefont {Botton}}, \bibinfo {author} {\bibfnamefont {S.~Y.}\ \bibnamefont {Savrasov}}, \bibinfo {author} {\bibfnamefont {C.}~\bibnamefont {Humphreys}},\ and\ \bibinfo {author} {\bibfnamefont {A.~P.}\ \bibnamefont {Sutton}},\ }\bibfield  {title} {\bibinfo {title} {Electron-energy-loss spectra and the structural stability of nickel oxide: An {LSDA + U} study},\ }\href@noop {} {\bibfield  {journal} {\bibinfo  {journal} {Physical Review B}\ }\textbf {\bibinfo {volume} {57}},\ \bibinfo {pages} {1505} (\bibinfo {year} {1998})}\BibitemShut {NoStop}%
\bibitem [{\citenamefont {Petukhov}\ \emph {et~al.}(2003)\citenamefont {Petukhov}, \citenamefont {Mazin}, \citenamefont {Chioncel},\ and\ \citenamefont {Lichtenstein}}]{lsda_2_petukhov2003correlated}%
  \BibitemOpen
  \bibfield  {author} {\bibinfo {author} {\bibfnamefont {A.}~\bibnamefont {Petukhov}}, \bibinfo {author} {\bibfnamefont {I.}~\bibnamefont {Mazin}}, \bibinfo {author} {\bibfnamefont {L.}~\bibnamefont {Chioncel}},\ and\ \bibinfo {author} {\bibfnamefont {A.}~\bibnamefont {Lichtenstein}},\ }\bibfield  {title} {\bibinfo {title} {Correlated metals and the {LSDA + U} method},\ }\href@noop {} {\bibfield  {journal} {\bibinfo  {journal} {Physical Review B}\ }\textbf {\bibinfo {volume} {67}},\ \bibinfo {pages} {153106} (\bibinfo {year} {2003})}\BibitemShut {NoStop}%
\bibitem [{\citenamefont {Mostofi}\ \emph {et~al.}(2014)\citenamefont {Mostofi}, \citenamefont {Yates}, \citenamefont {Pizzi}, \citenamefont {Lee}, \citenamefont {Souza}, \citenamefont {Vanderbilt},\ and\ \citenamefont {Marzari}}]{wannier90_mostofi2014updated}%
  \BibitemOpen
  \bibfield  {author} {\bibinfo {author} {\bibfnamefont {A.~A.}\ \bibnamefont {Mostofi}}, \bibinfo {author} {\bibfnamefont {J.~R.}\ \bibnamefont {Yates}}, \bibinfo {author} {\bibfnamefont {G.}~\bibnamefont {Pizzi}}, \bibinfo {author} {\bibfnamefont {Y.-S.}\ \bibnamefont {Lee}}, \bibinfo {author} {\bibfnamefont {I.}~\bibnamefont {Souza}}, \bibinfo {author} {\bibfnamefont {D.}~\bibnamefont {Vanderbilt}},\ and\ \bibinfo {author} {\bibfnamefont {N.}~\bibnamefont {Marzari}},\ }\bibfield  {title} {\bibinfo {title} {An updated version of {Wannier90}: A tool for obtaining maximally-localised wannier functions},\ }\href@noop {} {\bibfield  {journal} {\bibinfo  {journal} {Computer Physics Communications}\ }\textbf {\bibinfo {volume} {185}},\ \bibinfo {pages} {2309} (\bibinfo {year} {2014})}\BibitemShut {NoStop}%
\bibitem [{\citenamefont {Wu}\ \emph {et~al.}(2018)\citenamefont {Wu}, \citenamefont {Zhang}, \citenamefont {Song}, \citenamefont {Troyer},\ and\ \citenamefont {Soluyanov}}]{WannierTools_wu2018wanniertools}%
  \BibitemOpen
  \bibfield  {author} {\bibinfo {author} {\bibfnamefont {Q.}~\bibnamefont {Wu}}, \bibinfo {author} {\bibfnamefont {S.}~\bibnamefont {Zhang}}, \bibinfo {author} {\bibfnamefont {H.-F.}\ \bibnamefont {Song}}, \bibinfo {author} {\bibfnamefont {M.}~\bibnamefont {Troyer}},\ and\ \bibinfo {author} {\bibfnamefont {A.~A.}\ \bibnamefont {Soluyanov}},\ }\bibfield  {title} {\bibinfo {title} {Wanniertools: An open-source software package for novel topological materials},\ }\href@noop {} {\bibfield  {journal} {\bibinfo  {journal} {Computer Physics Communications}\ }\textbf {\bibinfo {volume} {224}},\ \bibinfo {pages} {405} (\bibinfo {year} {2018})}\BibitemShut {NoStop}%
\bibitem [{\citenamefont {Liechtenstein}\ \emph {et~al.}(1987)\citenamefont {Liechtenstein}, \citenamefont {Katsnelson}, \citenamefont {Antropov},\ and\ \citenamefont {Gubanov}}]{Liechtenstein1987}%
  \BibitemOpen
  \bibfield  {author} {\bibinfo {author} {\bibfnamefont {A.~I.}\ \bibnamefont {Liechtenstein}}, \bibinfo {author} {\bibfnamefont {M.~I.}\ \bibnamefont {Katsnelson}}, \bibinfo {author} {\bibfnamefont {V.~P.}\ \bibnamefont {Antropov}},\ and\ \bibinfo {author} {\bibfnamefont {V.~A.}\ \bibnamefont {Gubanov}},\ }\bibfield  {title} {\bibinfo {title} {Local spin density functional approach to the theory of exchange interactions in ferromagnetic metals and alloys},\ }\href@noop {} {\bibfield  {journal} {\bibinfo  {journal} {J. Magn. Magn. Mater.}\ }\textbf {\bibinfo {volume} {67}},\ \bibinfo {pages} {65 } (\bibinfo {year} {1987})}\BibitemShut {NoStop}%
\bibitem [{\citenamefont {Hoffmann}\ \emph {et~al.}(2020)\citenamefont {Hoffmann}, \citenamefont {Ernst}, \citenamefont {Hergert}, \citenamefont {Antonov}, \citenamefont {Adeagbo}, \citenamefont {Geilhufe},\ and\ \citenamefont {{Ben Hamed}}}]{Hoffmann2020}%
  \BibitemOpen
  \bibfield  {author} {\bibinfo {author} {\bibfnamefont {M.}~\bibnamefont {Hoffmann}}, \bibinfo {author} {\bibfnamefont {A.}~\bibnamefont {Ernst}}, \bibinfo {author} {\bibfnamefont {W.}~\bibnamefont {Hergert}}, \bibinfo {author} {\bibfnamefont {V.~N.}\ \bibnamefont {Antonov}}, \bibinfo {author} {\bibfnamefont {W.~A.}\ \bibnamefont {Adeagbo}}, \bibinfo {author} {\bibfnamefont {M.~R.}\ \bibnamefont {Geilhufe}},\ and\ \bibinfo {author} {\bibfnamefont {H.}~\bibnamefont {{Ben Hamed}}},\ }\bibfield  {title} {\bibinfo {title} {{Magnetic and Electronic Properties of Complex Oxides from First-Principles}},\ }\href@noop {} {\bibfield  {journal} {\bibinfo  {journal} {Phys. Status Solidi B}\ }\textbf {\bibinfo {volume} {257}},\ \bibinfo {pages} {1900671} (\bibinfo {year} {2020})}\BibitemShut {NoStop}%
\bibitem [{\citenamefont {Gyorffy}\ \emph {et~al.}(1985)\citenamefont {Gyorffy}, \citenamefont {Pindor}, \citenamefont {Staunton}, \citenamefont {Stocks},\ and\ \citenamefont {Winter}}]{Gyorffy1985}%
  \BibitemOpen
  \bibfield  {author} {\bibinfo {author} {\bibfnamefont {B.~L.}\ \bibnamefont {Gyorffy}}, \bibinfo {author} {\bibfnamefont {A.~J.}\ \bibnamefont {Pindor}}, \bibinfo {author} {\bibfnamefont {J.}~\bibnamefont {Staunton}}, \bibinfo {author} {\bibfnamefont {G.~M.}\ \bibnamefont {Stocks}},\ and\ \bibinfo {author} {\bibfnamefont {H.}~\bibnamefont {Winter}},\ }\bibfield  {title} {\bibinfo {title} {A first-principles theory of ferromagnetic phase transitions in metals},\ }\href@noop {} {\bibfield  {journal} {\bibinfo  {journal} {Journal of Physics F: Metal Physics}\ }\textbf {\bibinfo {volume} {15}},\ \bibinfo {pages} {1337} (\bibinfo {year} {1985})}\BibitemShut {NoStop}%
\bibitem [{\citenamefont {Staunton}\ \emph {et~al.}(1985)\citenamefont {Staunton}, \citenamefont {Gyorffy}, \citenamefont {Pindor}, \citenamefont {Stocks},\ and\ \citenamefont {Winter}}]{Staunton1985}%
  \BibitemOpen
  \bibfield  {author} {\bibinfo {author} {\bibfnamefont {J.~B.}\ \bibnamefont {Staunton}}, \bibinfo {author} {\bibfnamefont {B.~L.}\ \bibnamefont {Gyorffy}}, \bibinfo {author} {\bibfnamefont {A.~J.}\ \bibnamefont {Pindor}}, \bibinfo {author} {\bibfnamefont {G.~M.}\ \bibnamefont {Stocks}},\ and\ \bibinfo {author} {\bibfnamefont {H.}~\bibnamefont {Winter}},\ }\bibfield  {title} {\bibinfo {title} {Electronic structure of metallic ferromagnets above the curie temperature},\ }\href@noop {} {\bibfield  {journal} {\bibinfo  {journal} {Journal of Physics F: Metal Physics}\ }\textbf {\bibinfo {volume} {15}},\ \bibinfo {pages} {1387} (\bibinfo {year} {1985})}\BibitemShut {NoStop}%
\bibitem [{\citenamefont {Giri}\ \emph {et~al.}(2020)\citenamefont {Giri}, \citenamefont {Mallick}, \citenamefont {Singh}, \citenamefont {Madduri}, \citenamefont {Damay}, \citenamefont {Alam},\ and\ \citenamefont {Nayak}}]{heusler_mn2pt_giri2020robust}%
  \BibitemOpen
  \bibfield  {author} {\bibinfo {author} {\bibfnamefont {B.}~\bibnamefont {Giri}}, \bibinfo {author} {\bibfnamefont {A.~I.}\ \bibnamefont {Mallick}}, \bibinfo {author} {\bibfnamefont {C.}~\bibnamefont {Singh}}, \bibinfo {author} {\bibfnamefont {P.~P.}\ \bibnamefont {Madduri}}, \bibinfo {author} {\bibfnamefont {F.}~\bibnamefont {Damay}}, \bibinfo {author} {\bibfnamefont {A.}~\bibnamefont {Alam}},\ and\ \bibinfo {author} {\bibfnamefont {A.~K.}\ \bibnamefont {Nayak}},\ }\bibfield  {title} {\bibinfo {title} {Robust topological hall effect driven by tunable noncoplanar magnetic state in mn-pt-in inverse tetragonal heusler alloys},\ }\href@noop {} {\bibfield  {journal} {\bibinfo  {journal} {Physical Review B}\ }\textbf {\bibinfo {volume} {102}},\ \bibinfo {pages} {014449} (\bibinfo {year} {2020})}\BibitemShut {NoStop}%
\bibitem [{\citenamefont {Xu}\ \emph {et~al.}(2020)\citenamefont {Xu}, \citenamefont {Elcoro}, \citenamefont {Song}, \citenamefont {Wieder}, \citenamefont {Vergniory}, \citenamefont {Regnault}, \citenamefont {Chen}, \citenamefont {Felser},\ and\ \citenamefont {Bernevig}}]{mvasp2trac1_xu2020high}%
  \BibitemOpen
  \bibfield  {author} {\bibinfo {author} {\bibfnamefont {Y.}~\bibnamefont {Xu}}, \bibinfo {author} {\bibfnamefont {L.}~\bibnamefont {Elcoro}}, \bibinfo {author} {\bibfnamefont {Z.-D.}\ \bibnamefont {Song}}, \bibinfo {author} {\bibfnamefont {B.~J.}\ \bibnamefont {Wieder}}, \bibinfo {author} {\bibfnamefont {M.}~\bibnamefont {Vergniory}}, \bibinfo {author} {\bibfnamefont {N.}~\bibnamefont {Regnault}}, \bibinfo {author} {\bibfnamefont {Y.}~\bibnamefont {Chen}}, \bibinfo {author} {\bibfnamefont {C.}~\bibnamefont {Felser}},\ and\ \bibinfo {author} {\bibfnamefont {B.~A.}\ \bibnamefont {Bernevig}},\ }\bibfield  {title} {\bibinfo {title} {High-throughput calculations of magnetic topological materials},\ }\href@noop {} {\bibfield  {journal} {\bibinfo  {journal} {Nature}\ }\textbf {\bibinfo {volume} {586}},\ \bibinfo {pages} {702} (\bibinfo {year} {2020})}\BibitemShut {NoStop}%
\bibitem [{\citenamefont {Elcoro}\ \emph {et~al.}(2021)\citenamefont {Elcoro}, \citenamefont {Wieder}, \citenamefont {Song}, \citenamefont {Xu}, \citenamefont {Bradlyn},\ and\ \citenamefont {Bernevig}}]{mvasp2trac2_elcoro2021magnetic}%
  \BibitemOpen
  \bibfield  {author} {\bibinfo {author} {\bibfnamefont {L.}~\bibnamefont {Elcoro}}, \bibinfo {author} {\bibfnamefont {B.~J.}\ \bibnamefont {Wieder}}, \bibinfo {author} {\bibfnamefont {Z.}~\bibnamefont {Song}}, \bibinfo {author} {\bibfnamefont {Y.}~\bibnamefont {Xu}}, \bibinfo {author} {\bibfnamefont {B.}~\bibnamefont {Bradlyn}},\ and\ \bibinfo {author} {\bibfnamefont {B.~A.}\ \bibnamefont {Bernevig}},\ }\bibfield  {title} {\bibinfo {title} {Magnetic topological quantum chemistry},\ }\href@noop {} {\bibfield  {journal} {\bibinfo  {journal} {Nature communications}\ }\textbf {\bibinfo {volume} {12}},\ \bibinfo {pages} {5965} (\bibinfo {year} {2021})}\BibitemShut {NoStop}%
\bibitem [{\citenamefont {Iraola}\ \emph {et~al.}(2022)\citenamefont {Iraola}, \citenamefont {Ma{\~n}es}, \citenamefont {Bradlyn}, \citenamefont {Horton}, \citenamefont {Neupert}, \citenamefont {Vergniory},\ and\ \citenamefont {Tsirkin}}]{iraola2022irrep}%
  \BibitemOpen
  \bibfield  {author} {\bibinfo {author} {\bibfnamefont {M.}~\bibnamefont {Iraola}}, \bibinfo {author} {\bibfnamefont {J.~L.}\ \bibnamefont {Ma{\~n}es}}, \bibinfo {author} {\bibfnamefont {B.}~\bibnamefont {Bradlyn}}, \bibinfo {author} {\bibfnamefont {M.~K.}\ \bibnamefont {Horton}}, \bibinfo {author} {\bibfnamefont {T.}~\bibnamefont {Neupert}}, \bibinfo {author} {\bibfnamefont {M.~G.}\ \bibnamefont {Vergniory}},\ and\ \bibinfo {author} {\bibfnamefont {S.~S.}\ \bibnamefont {Tsirkin}},\ }\bibfield  {title} {\bibinfo {title} {Irrep: symmetry eigenvalues and irreducible representations of ab initio band structures},\ }\href@noop {} {\bibfield  {journal} {\bibinfo  {journal} {Computer Physics Communications}\ }\textbf {\bibinfo {volume} {272}},\ \bibinfo {pages} {108226} (\bibinfo {year} {2022})}\BibitemShut {NoStop}%
\end{thebibliography}%

\end{document}